\documentclass[prl,aps,amssymb,nofootinbib,twocolumn,showpacs]{revtex4}
\usepackage{amsmath}
\usepackage{amssymb}
\usepackage{amsthm}
\usepackage{amsfonts}
\usepackage{algorithmic}
\usepackage{enumerate}
\usepackage{latexsym}
\usepackage[dvips]{graphicx}

\newcommand{\beq}{\begin{equation}}
\newcommand{\eneq}{\end{equation}}
\newcommand{\beqs}{\begin{equation*}}
\newcommand{\eneqs}{\end{equation*}}

\input{epsf}

\begin{document}

\tolerance 10000

\title{Trapping of Quantized Electrical Charge in Superfluid $^3$He-B via the
Electrodynamics of Spin-orbit Coupled Systems}

\author { Zaira Nazario$^\dagger$, Bart Leurs\S, David
I. Santiago$^{\dagger,\star}$ and Jan Zaanen$^\dagger$\S}

\affiliation{ $\dagger$ Department of Physics, Stanford
University,Stanford, California 94305 \\
$\star$ Gravity Probe B Relativity Mission, Stanford, California 94305
\\
\S Instituut Lorentz for Theoretical Physics, Leiden University,
Leiden, The Netherlands}

\begin{abstract}
\begin{center}

\parbox{14cm}{Exploiting analogies between spin-orbit coupled spin
superfluids and non-Abelian Yang-Mills theory, we argue that machines
can be built capable of trapping a quantized amount of electric linear
charge density, while the line charge quantum itself is surprisingly
large. The required conditions might well hold in superfluid $^3$He-B,
for which we propose an experimental realization of this phenomenon.}

\end{center}
\end{abstract}

\date{\today}

\maketitle

The subject of spintronics in semiconductors recently received new
impetus in the form of the spin Hall effect arising from spin-orbit
coupling\cite{spinhall1}. This is governed by an elegant macroscopic
transport equation,
    \begin{equation}
      j^a_i = \sigma_{S,H} \epsilon_{ial} E_l
      \label{spinhall}
    \end{equation}
where $\epsilon_{ial}$ is the 3-dimensional antisymmetric tensor,
$E_l$ the electrical field and $j^a_i$ the spin-current ($a$ internal
spin- and $i$ embedding space direction). Since both $j^a_i$ and $E_l$
are even under time reversal, the transport coefficient $\sigma_{S,H}$
is also even under time reversal, indicating that it is fundamentally
a dissipationless transport phenomenon.  According to the present
concensus, the spin-Hall effect of the semiconductors can be viewed as
a genuine quasi-classical transport phenomenon\cite{spinhall2}. This
has its drawbacks since Eq. (\ref{spinhall}) does not have truly
hydrodynamic status because classical spin currents are not conserved
in the presence of spin orbit coupling.

Is there another, in a way deeper meaning to Eq. (\ref{spinhall})?
Like matter fluids, also spin fluids can occur in quantum-coherent
incarnations as spin superfluids. The B superfluid phase of $^3$He is
of this kind as it breaks spin rotational
invariance\cite{leggett,volovik,volovik2}. In the presence of
spin-orbit coupling, these order-parameter theories take the form of
non-Abelian Yang-Mills theories with Higgs fields, albeit with the
specialty that the gauge fields occur in a `fixed-frame': these
actually are set by the physical electromagnetic fields.  The
spin-Hall equation Eq.(\ref{spinhall}) acquires now a most special
meaning: it is nothing else than the non-Abelian London equation `in
the fixed frame', the constituent equation catching the essence of the
quantum hydrodynamics associated with the non-Abelian generalizations
of the Meissner effect.

The relation between spin-orbit coupling and Yang-Mills gauge
structures in the fixed frame is rooted in the Pauli
equation\cite{gold}.  This was realized by Mineev and Volovik in the
context of $^3$He-B\cite{volovik}. Some of the physical ramifications
of these theoretical observations have been explored by Balatsky and
Altshuler\cite{ba} who made predictions of Aharonov-Casher\cite{ac}
spin-orbit phase interferences in $^3$He-A$_1$ driven by a fixed
external electric field. We expand on their work and obtain the other
side of the coin. We predict that the persistent currents of a spin
superfluid like $^3$He-B are capable to trap a quantized charge.
    \begin{figure}[ht]
      \centering
      \rotatebox{0}{
    \resizebox{5.3cm}{!}{%
      \includegraphics*{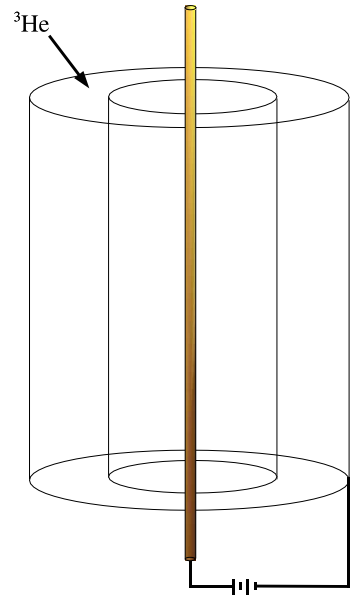}}}
      \caption{A superfluid $^3$He-B container acts as a capacitor capable
      of trapping a quantized electrical line charge density via the electric
      field generated by persistent spin Hall currents. This is the analog
      of magnetic flux trapping in superconductors by persistent
      charge supercurrents.}
    \end{figure}

We follow the experimental set-up proposed by Balatsky and
Altshuler\cite{ba} and consider a cylindrical glass (or plastic)
container of inner radius $R_1$ and outer radius $R_2$ filled with
$^3$He-B, threaded by a metal wire of radius $a$ (Fig. 1). The outside
cylindrical surface is plated with a grounded metal. The wire and the
outer metal cylinder form a capacitor and a bias is switched to charge
the wire while $^3$He is still in the normal phase. $^3$He is cooled
through its superfluid transition. Then the bias source is quickly
removed. Upon attempting to discharge the wire by touching it to
ground, it will not discharge: the charge per unit length left in the
wire will be $N\lambda_0$, where $N$ is an integer, while the `fat'
elementary linear charge density quantum amounts to,
    \beq
      \lambda_0 = \left(\frac{\hbar c^2}{4\mu} \right) \sim 3.5 \times
      10^{-7} Coulomb/meter \; .
      \label{chargequant}
    \eneq
$\mu$ is the magnetic moment of $^3$He atoms and $c$ is the speed of
light. A quantized charge is trapped!  The only way to get the charge
out is to heat the Helium and right at the superfluid transition, the
quantized charge will be released. This quantization is due to
the constructive interference of the Aharonov-Casher\cite{ba,ac} phase
of the $^3$He-B order parameter, in close analogy to magnetic
flux-quantization and -trapping in normal superconductors.

We first consider a hypothetical pure spin superfluid with an $SU(2)$
spin degree of freedom -- the real life case of $^3$He-B is
qualitatively the same but complexer in detail due to the mixed
spin-orbital nature of its order parameter. The starting
point is the Pauli equation, containing the leading relativistic
corrections, which can be casted in the form of a $SU(2)$ Yang-Mills
Schr\"odinger equation\cite{gold,volovik},
    \beq
      \label{pauli}
      i\hbar D_0\psi = -\frac{\hbar^2}{2m} D_i^2\psi + V \psi
    \eneq
with $D_i=\partial_i - i \frac{q}{\hbar c} A_i^a \frac{\tau^a}{2}$,
$D_0=\partial_0 + i \frac{q}{\hbar} A_0^a \frac{\tau^a}{2}$ where q is the
magnetic moment of particle. This is
true as long as one takes a particular `gauge-fix' which actually
amounts to identifying the Yang-Mills gauge fields with the physical
electrical (E) and magnetic (B) fields,
    \beq
      A^a_i = \epsilon_{ial} E_l \; , \qquad A^a_0 = B^a
      \label{YMfix}
    \eneq
where $a$ and $0$ denote space and time directions, respectively. The Maxwell
equation $\nabla\times\vec E + \frac{\partial \vec B}{\partial t} = 0$ implies
$\partial^\mu A_\mu^a= 0$, thus forcing the Lorentz gauge condition on the
non-Abelian fields.

The important point is that spin-orbit coupling leads to parallel spin
transport governed by the Yang-Mills connection. One easily infers the
Ginzburg-Landau-Wilson action describing the spin superfluid order
parameter dynamics. The non-Abelian order parameter is $\Psi = g |
\Psi |$\cite{us} with $g= \exp{(i\varphi^a \tau^a/2)}$ and $\varphi_a$
the non-Abelian phase. Focussing on the gradient terms,
    \beq
      F = |(\partial_i - i \frac{q}{\hbar c} A^{a}_i \tau_a) \Psi^a|^2 +
      \dots = \rho_s (\omega^a_i - \frac{q}{\hbar c}A^a_i \tau_a)^2 + \dots
      \label{freeen}
    \eneq
with non-Abelian phase velocity
    \beq
      \label{navel}
      \omega^a_i \equiv -i Tr[g^{-1} \partial_i g \tau^a] = (1/2)
      \epsilon^{abc} \tilde R_{bd} \partial_i \tilde R_{cd}
    \eneq
where $\tilde R^a_{b}(\vec \varphi) \frac{\tau^b}{2} =
e^{-i\varphi^a\tau^a/2} \frac{\tau^a}{2} e^{i\varphi^a\tau^a/2}$. This
`spin' phase velocity is the non-Abelian generalization of the
superfluid velocity associated with Abelian superfluids and
superconductors. The fact that in the latter the velocity can be
written as the gradient of a scalar function just implies that
contrary to classical fluids, quantum hydrodynamics is
irrotational. For the non-Abelian fluid this is not true generally as
it obeys a Mermin-Ho type relation, $\nabla \times \vec{\omega}^a =
\epsilon_{abc} \vec{\omega}^b \times\vec{\omega}^c$.

Lacking genuine conservation laws, the non-Abelian classical fluid is
not governed by a universal hydrodynamic behavior. However, when the
fluid turns quantum coherent this changes drastically:
Eq. (\ref{freeen}) is an order parameter theory as well behaved as any
other and there is a true sense of quantum hydrodynamics. After all,
in the full Yang-Mills theory, the phase currents are supposedly
responsible for giving mass to nature through the Higgs mechanism. The
non-Abelian London equation responsible for the Higgs mechanism is
obtained by varying to $\vec{\omega}^a$ in Eq. (\ref{freeen}):
$\vec{\omega}^a = \frac{q}{\hbar c} \vec{A}^a$.  In the spin-orbit
coupled spin-superfluid $A_i^a = \epsilon_{ial} E_l$ (the fixed frame,
Eq. \ref{YMfix}) and we directly recover the spin Hall relation,
Eq.(\ref{spinhall})!  This reveals the great depth of the spin-Hall
relation, dealing with quantum-coherent matter: the only meaningful
currents in the spin superfluid are spin-Hall currents!

The ramifications of superfluid hydrodynamics have to do with
quantized topological numbers. In the full $SU(2)$ scalar Higgs
Yang-Mills theory these are the 't Hooft-Polyakov
monopoles\cite{thooft}, point-like analogs of the magnetic flux
lines of normal superconductors, characterized by $\omega^a_i = A^a_i
= \epsilon_{ail} x_l / r^2$. The `fixed frame' point charge electric field is
$A^a_i \sim \epsilon_{ail} x_l / r^3$. In contrast to the 't
Hooft-Polyakov monopole, it corresponds to a field strength with a wrong
radial dependence for a nontrivial topology. On the
other hand, the electric field associated with the wire configuration
in Fig.1 is $E_i = 2\lambda \frac{x_{i\perp}}{r^2}$, $E_3=0$, where
$x_{i\perp}$ is $x$ or $y$ and $r$ is the radial direction in the $xy$
plane. The non-zero fixed frame gauge fields become,
    \beq
      \label{wireA}
      A_1^3 = -2\lambda\frac{x_2}{r^2},\ A_2^3 =
      2\lambda\frac{x_1}{r^2},\ A_3^2 = -A_2^3,\ A_3^1 = -A_1^3.
    \eneq
These gauge field configurations are a special case of the general
$SU(2)$ BPST-type line textures discovered by Witten in the
1970's\cite{witten}, obtained by choosing the gauge fix
$A_0$$=$$A_1$$=$$\phi_1$$=$$0$ and $\phi_2$$=$$\lambda$$-1$. The
relevant symmetry for the topological stability is $U(1)$, allowing
for stable quantized vortices. The full order parameter $\exp{(i\tau
_a A_i^a dx^i)}$$\in$$SU(2)$$\simeq$$S^3$, but with the 'wired in'
topology it becomes $U(1)$. Indeed, if we construct a vector $\alpha$
with elements $\alpha_k$$=$$\epsilon_{ijk}A_{ij}$, we obtain a
$U(1)$-subgroup of the rotations $SO(3)\ni \exp{(i\tau ^i\alpha_i)} =
\exp(\frac{2\lambda i}{r^2}(x_1,x_2)\cdot(\tau_1,\tau_2))$, i.e.,
$U(1)$-rotations about an axis.

In standard Higgs-Yang-Mills theory the above would be a sufficient
condition for spin-superfluid vorticity quantization, which leads to
line charge trapping. Here we are dealing with physical fields playing
the role of gauge fields. Does that make a difference?  Fibre bundles
are topologically classified by Chern classes, which are properties of
the gauge group, i.e., transition functions on the bundle, and not of
the special connection (gauge field) or standard fibre (Higgs field)
chosen. Our fixed frame fields just correspond to a particular
gauge. Regardless of the fact that they are actually physical fields,
they `impose' the topological invariants on the matter sector as
well. Hence, spin superfluid vorticity is quantized. At least for
topological purposes one can rely on the spin-Hall relation
(\ref{spinhall}) and infer that the spin superflow is of the
Abelianized Aharonov-Casher form\cite{ac}, $v^3_1 = - 2 \lambda
\frac{q}{\hbar c} \frac{x_2}{r^2}; \; v^3_2 = 2 \lambda \frac{q}{\hbar
c} \frac{x_1}{r^2}$ and it is a straightfoward excercise to show that
this leads to the quantization condition Eq. (\ref{chargequant}).
This completes our proof of principle.

The $^3$He superfluids are the physical systems approaching the above
mathematical ideal most closely. The order parameters in the $A$ and
$B$ phase are of a mixed spin-orbital nature which complicates matters
but not to the extent that the essential physics fails -- for example,
it is well known in the $^3$He literature\cite{volovik} that the B
phase can support spins supercurrents as it breaks spin rotational
invariance\cite{leggett,volovik,volovik2}. We will analyze this in
great detail elsewhere\cite{us} and let us present here a sketch
addressing a simplified version of the B phase. Let us first summarize
the basics, due to Mineev and Volovik\cite{volovik}.  The $^3$He-B
order parameter is described by, \cite{leggett,volovik,volovik2}
    \beq
      A^B_{\alpha j} = \Delta_B e^{i\Phi} R_{\alpha j}
      \label{heBorder}
    \eneq
where $\Delta_B$ is the amplitude, $\Phi$ the phase associated with
number, and $R_{\alpha j}(\hat n,\theta)$ is a matrix associated with
the spin- ($S, \alpha$) and orbital ($L, j$) degrees of freedom such
that it describes a rotation by an angle $\theta$ about the arbitrary
ordering direction $\hat n\propto\langle\vec L\times\vec S
\rangle$. This describes the breaking of spin- and orbital rotational
invariance separately while the total angular invariance $L + S$ is
unbroken. The order parameter matrix is $R_{\alpha j} = R^S_{\alpha
i}R^L_{ij}$ where $R^S$ and $R^L$ describe pure spin- and orbital
rotations, respectively. 

>From the comparison with Eq. (\ref{navel}) it follows that one
can identify a quantity which is uniquely associated with the
spin-only  phase velocity field $\omega_{\alpha i}$
and magnetization density $\omega_{\alpha}$ \cite{volovik},
    \begin{eqnarray}
      \omega_{\alpha i} & = & \frac{1}{2}
      \epsilon_{\alpha\beta\gamma}R_{\beta j} \partial_i R_{\gamma j}\; ,
      \nonumber \\
      \omega_\alpha & = & \frac{1}{2}
      \epsilon_{\alpha\beta\gamma}R_{\beta j} \partial_0 R_{\gamma j}\; .
      \label{omegai}
    \end{eqnarray}
The central observation of Mineev and Volovik\cite{volovik} is that
electrical fields as mediated by spin-orbit coupling enter the
$^3$He-B superfluid hydrodynamics exclusive by their coupling to the
above spin superfluid velocity field. This is exactly equivalent to
how spin orbit coupling enters the ideal spin superfluid, Eq.
(\ref{freeen}). The $^3$He-B order parameter ``Higgs'' Lagrangian is
    \beq
      \label{laghe}
      L(R_{\alpha j},A_0,\vec A) = L_{\text{kin}}(R_{\alpha j}) +
      F_{\text{grad}}(R_{\alpha j}) + \frac{1}{8\pi}\left(E^2 -
      B^2\right)
    \eneq
    \begin{align}
      &\nonumber\text{where } L_{\text{kin}}(R_{\alpha j}) =
      -\frac{n_s\hbar^2}{2mc_s^2} \left(\vec\omega^2 + \frac{4\mu}
      {\hbar}\vec\omega\cdot\vec B \right) \\
      &\nonumber F_{\text{grad}}(R_{\alpha, j}) = \frac{1}{2}
      \rho_{\alpha i, \beta j}\left(\omega_{\alpha i}\omega_{\beta j}
      - \frac{4\mu} {\hbar c^2}\omega_{\alpha i}\epsilon_{\alpha ik}
      E_k\right) \\
      &\nonumber \rho_{\alpha i, \beta j} =
      \frac{n_s\hbar^2}{mc_s^2}[\tilde
      c_\parallel^2\delta_{\alpha\beta}\delta_{ij} - (\tilde
      c_\parallel^2 - \tilde c_\perp^2)(R_{\alpha i}R_{\beta j} +
      R_{\alpha j}R_{\beta i})] \; .
    \end{align}
with $n_s$ the superfluid density, $m$ the mass of the $^3$He atoms,
$\tilde c_\parallel^2 = c_\perp^2$ and $\tilde c_\perp^2 =
\left(1/2\right)\left(c_\perp^2 + c_\parallel^2\right)$, while
$c_\parallel$, $c_\perp$ and $c_s$ are the longitudinal-, transversal-
and average spin wave velocities.

As we show elsewhere\cite{us}, the spin wave velocity anisotropy
cannot change the topological structure of the system. Hence we will
neglect it supposing $^3$He-B to be spin isotropic, $\rho_{\alpha
i,\beta j} = \frac{n_s\hbar^2}{m}
\delta_{\alpha\beta}\delta_{ij}$. Let us now depart from Mineev and
Volovik, to find out the equations of motions associated with the
Lagrangian Eq (\ref{laghe}). By varying it with respect to scalar- and
vector potentials respectively, we obtain the pair of Maxwell
equations,
   \begin{eqnarray}
      \partial_kE_k & = & 4\pi \partial_k\left(
      \frac{2n_s\mu}{m c^2}\epsilon_{\alpha
      ik}\omega_{\alpha i} \right)
      \label{efieldhe} \\
      \left(\nabla\times\vec B\right)_\alpha & = & -
    4\pi\left(\nabla\times
    \frac{2n_s\mu\hbar}{m c_s^2}\omega_\alpha \right) +
    \partial_0 D_\alpha
      \label{bfieldhe}
    \end{eqnarray}

These are just the usual Maxwell equations associated with
 the electric displacement $\nabla\cdot\vec D = 0$ where
$\vec D = \vec E + 4\pi \vec P$ with
    \beq
      P_k = - \frac{2n_s\mu}{m c^2}\epsilon_{\alpha
      ik}\omega_{\alpha i} \; .
    \eneq

The spin current acts as a polarization for the medium because, upon
Lorentz transforming, the magnetic field in the frame of the moving
spin turns into the lab frame in an electric field.  This is of course
well known, but different from the semiconductors\cite{rashba} $^3$He
is an electrically very quiet environment where the only other source
of electrical fields comes from deformation of the electronic shells
of the $^3$He atoms caused by gradients of the order
parameter\cite{volovik}. It appears to us that by electrical
measurements much can be learned about spin superflow in $^3$He.

Let us now focus on the equations of motion of the spin superfluid
associated with Eq. (\ref{laghe}),
    \begin{align}
      \begin{aligned}
    \label{eqmot}
    0 &= \partial_0 \left[ \frac{1}{2}\epsilon_{\alpha\beta\gamma}
    R_{\gamma j} \frac{n_s\hbar^2}{mc_s^2} \left(\omega_\alpha +
    \frac{2\mu}{\hbar} B_\alpha\right) \right] \\
    &+ \partial_i \left[ - \frac{1}{2}\epsilon_{\alpha\beta\gamma}
    R_{\gamma j} \frac{n_s\hbar^2}{m} \left(\omega_{\alpha i} -
    \frac{2\mu}{\hbar c^2} \epsilon_{\alpha ik} E_k\right)
    \right] \\
    &+\frac{1}{2}\epsilon_{\alpha\beta\gamma}(\partial_0 R_{\gamma
    j}) \frac{n_s\hbar^2}{mc_s^2} \left(\omega_\alpha + \frac{2\mu}
    {\hbar} B_\alpha\right) \\
    &- \frac{1}{2}\epsilon_{\alpha\beta\gamma}(\partial_i
    R_{\gamma j}) \frac{n_s\hbar^2}{m} \left(\omega_{\alpha i} -
    \frac{2\mu}{\hbar c^2} \epsilon_{\alpha ik} E_k\right) \;
    .
      \end{aligned}
    \end{align}
This equation is quite a bit more involved than the simple spin-Hall
relation Eq. (\ref{spinhall}) of the $SU(2)$ superfluid. This reflects
the fundamental difference between B-phase order parameter and a pure
spin condensate: in the B-phase, the order parameter is the `relative
spin-orbital' $SO(3)$ while only the spin-transport is `gauged' by
spin-orbital coupling. We nevertheless managed to find a solution of
Eq.'s (\ref{eqmot},\ref{efieldhe}) for the geometry in Fig. 1, demonstrating
that despite these complications the B-phase does quantize the line
charge density in exactly the same way as the $SU(2)$ superfluid.

This solution is as follows. As before, the electrical field will be
radial ($\vec E=$$\frac{2\lambda}{r}$$\hat r$) and we can invoke the
same argument as we did for the $SU(2)$ superfluid, which becomes even
simpler since the symmetry of the Higgs field is $SO(3)$. The
cylindrical symmetry of the electrical field may now be interpreted as
an $SO(2) = U(1)$ gauge field directly. It carries an integer
topological charge, and as before the Higgs field will inherit this
charge as well. Hence, topologically the spin sector is vortex like.
As we will discuss in a moment, the electrical field strength is not
affected by the presence of the Helium and it follows that the spin
current is, \beq
      \label{hespcu}
      \vec\omega_z = -\frac{4\lambda\mu}{\hbar c^2r}\hat\varphi
      \;.
    \eneq
The charge per unit length, $\lambda$, in the wire is given
by the potential difference $V_\text{battery} = 2\lambda\ln
\frac{R_2}{a}$.

We have now to reconcile this constraint coming from the gauge side
with the relative spin-orbital nature of the order parameter. To
satisfy this constraint, the spins should be in the z-direction always
and this implies that the spin rotation matrix $R^S_{\alpha\beta}$
should be taken to be the identity matrix
$\delta_{\alpha\beta}$. Since the superfluid is flowing in the
azimuthal direction, and since the spin part is diagonal, we have to
take for the orbital rotation matrix $R^L_{ij}$ and the Helium order
parameter $R_{\alpha j}$, 
    \beq 
      R_{\alpha j} = \left (
      \begin{matrix}
    \cos\theta & -\sin\theta & 0 \\
    \sin\theta & \cos\theta & 0 \\
    0 & 0 & 1
      \end{matrix}
      \right) \; ,
    \eneq
in terms of the relative angle $\theta$. Using the above in the definition 
of the spin current (Eq. \ref{omegai}) we find: $\vec\omega_z =
-\nabla\theta$. Remarkably, the spin current is now also associated with
the gradient of the relative angle, a quantity associated with the genuine
B-phase order parameter! The single valuedness of the order parameter
implies the  quantization condition upon going around the cylinder ($N$
is the winding number),
    \beq
      \oint\vec\omega_z\cdot d\vec l = 2\pi N \; ,
      \label{quantcond}
    \eneq
Upon integrating Eq. (\ref{hespcu}) we obtain $\frac{4\lambda\mu} {\hbar
c^2} 2\pi$. Together with Eq. (\ref{quantcond}) this implies that
the wire will be left with a quantized line charge,
    \beq
      \lambda' = N \left( \frac{\hbar c^2}{4\mu} \right) \; .
      \label{quanthe}
    \eneq

Therefore, if we measure the potential difference between the wire and
the outside cylinder, it will be nonzero even after shorting out. It
will be
    \beq
      V = 2N \left( \frac{\hbar c^2}{4\mu} \right) \ln\frac{R_2}{a}
      \; .
      \label{volthe}
    \eneq
This potential is much larger than atomic potentials because of the
smallness of the spin orbit coupling constant $\mu/\hbar c^2$ to which
it is inversely proportional. We notice that this requires that the
dimensions of the vessel should be less than the dipolar length
$\simeq 10 \mu$m because at larger distances a `soliton tail' develops
because $\theta$ pins to the Leggett angle\cite{kondo}.

In conclusion, for quite non-trivial reasons $^3$He B-phase mimics
perfectly the charge quantization effect of the idealized spin-orbit
coupled spin superfluid, and we leave it to the experts to find out if
this device is technically feasible. To stress how remarkable this
effect is,let us consider what actually happens with the electrical
field in the presence of the spin-fluid or B-phase.  In analogy with
normal superconductors, the spin Hall- (Eq. \ref{spinhall}) and
Maxwell (Eq. \ref{efieldhe}) equations take the role of the London-
and magnetic ($\epsilon_{abc} \partial_b B_c \sim J_a$) Maxwell
equations. However, instead of the Meissner effect, if follows that $(
1 - const.) \partial_k E_k = 0$ showing that the electrical field is
not at all affected by the presence of spin-matter! This actually
makes sense: the appropriate analogy is that the electrical charge in
the wire takes the role of magnetic flux, and the electrical field
that of the vector potential, and we encounter a quite `material'
version of the ghostly action on a distance discovered by Aharonov and
Bohm. It is `material' at least in the sense that the effect lowers
the free energy. For a constant electrical field it follows from
Eq. (\ref{laghe}) that the spin currents lower the total energy by
$\Delta E = 2n_s\mu^2/mc^4)E^2$.

\noindent {\bf Acknowledgements:} We acknowledge helpful discussions
with G.E. Volovik, A.V.  Balatsky, S.C. Zhang, A. Gernevik,
B. I. Halperin, M. Esole and E.I. Rashba. B.L. and J. Z. were
supported by the Dutch Science Foundation NWO/FOM. J. Z. acknowledges
support by the Fulbright Foundation in the form of a senior
fellowship. Z. N. was supported by the Ford Foundation and by the
School of Humanities and Science at Stanford University. D. I. S. was
supported by NASA Grant NAS 8-39225 to Gravity Probe B.

\end{document}